\documentclass[prl,a4paper,twocolumn,nofootinbib,superscriptaddress]{revtex4-1}

\usepackage[utf8]{inputenc}
\usepackage{amssymb}
\usepackage{amsmath}
\usepackage{graphicx}
\usepackage{psfrag}
\usepackage{latexsym}
\usepackage{hyperref}

\newcommand{\qnm}{\tilde{\Pi}}
\newcommand{\eq}{\begin{equation}}
\newcommand{\eqx}{\end{equation}}
\newcommand{\eqn}{\begin{eqnarray}}
\newcommand{\eqnx}{\end{eqnarray}}

\newcommand{\f}[2]{\frac{#1}{#2}}
\newcommand{\rf}[1]{(\ref{#1})}
\newcommand{\nn}{{\mathcal N}}

\newcommand{\D}{{\mathcal D}}

\begin{document}

\title{Coupling hydrodynamics to nonequilibrium degrees of freedom in strongly
  interacting quark-gluon plasma}

\author{Michal P. Heller}
\affiliation{\it Instituut voor Theoretische Fysica, Universiteit van
  Amsterdam, Science Park 904, 1090 GL Amsterdam, The Netherlands} 
\affiliation{\it Perimeter Institute for Theoretical Physics, Waterloo, Ontario N2L 2Y5, Canada} 
\affiliation{National Centre for Nuclear Research,
  Ho{\.z}a 69, 00-681 Warsaw, Poland}

\author{Romuald A. Janik}
\affiliation{Institute of Physics,
Jagiellonian University, Reymonta 4, 30-059 Krak\'ow, Poland}

\author{Micha\l\ Spali\'nski}
\affiliation{National Centre for Nuclear Research,
  Ho{\.z}a 69, 00-681 Warsaw, Poland}
\affiliation{Physics Department, University of Bia{\l}ystok, Lipowa 41, 15-424 Bia{\l}ystok, Poland}

\author{Przemys{\l}aw Witaszczyk}
\affiliation{Institute of Physics,
Jagiellonian University, Reymonta 4, 30-059 Krak\'ow, Poland}

\begin{abstract}

Relativistic hydrodynamics simulations of quark-gluon plasma play a pivotal
role in our understanding of heavy ion collisions at RHIC and LHC. They are
based on a phenomenological description due to M{\"u}ller, Israel, Stewart
(MIS) and others, which incorporates viscous effects and ensures a well-posed
initial value problem. Focusing on the case of conformal plasma 
we propose a generalization which includes, in addition,
the dynamics of the least damped far-from-equilibrium degree of freedom found
in strongly coupled plasmas through the AdS/CFT correspondence. We formulate
new evolution equations for general flows and then test them in the case of
${\cal N}=4$ super Yang-Mills plasma by comparing their solutions alongside
solutions of MIS theory with numerical computations of isotropization and
boost-invariant flow based on holography. In these tests the new equations
reproduce the results of MIS theory when initialized close to the hydrodynamic
stage of evolution, but give a more accurate description of the dynamics when
initial conditions are set in the pre-equilibrium regime.

\end{abstract}

\maketitle

\section{Introduction}

The successful phenomenological description of soft observables in heavy ion
collisions at RHIC and LHC asserts that the quark-gluon plasma phase is formed
and in less than one Fermi after the collision subsequent evolution till
hadronization is governed by hydrodynamic expansion with a very small shear
viscosity \cite{Heinz:2004pj,Wiedemann:2012py}. Finding an explanation for the
emergence of this collective behavior under experimentally viable conditions
based on the microscopic theory, QCD, poses a timely theoretical challenge. In
consequence, much attention has recently been devoted to the studies of
equilibration processes of non-Abelian gauge fields in a few known tractable
situations, such as at strong coupling using holography and a dual
gravitational description.

Within this approach it has been shown that viscous hydrodynamics can work
remarkably well already after a time of order of the inverse of the local effective
temperature\footnote{The hydrodynamization in $1/T$ is
  phenomenologically attractive, as ballpark quantities characterizing
  initialization of hydrodynamics codes, $\tau = 0.5 \, \mathrm{Fermi}$ and $T
  = 500 \, \mathrm{MeV}$, obey $\tau = {\cal O}(1) / T$.} despite significant
pressure anisotropy in the local rest frame
\cite{Chesler:2009cy,Chesler:2010bi,Heller:2011ju,Casalderrey-Solana:2013aba}. This
finding suggests that the applicability of hydrodynamics is not limited by the
size of gradient corrections to the perfect fluid stress tensor, but rather by
the presence of degrees of freedom not described by hydrodynamics. This
implies that any phenomenological attempts to capture features of
pre-equilibrium dynamics in heavy ion collisions need to incorporate effects
of these degrees of freedom.

The holographic AdS/CFT description of ${\cal N}=4$ supersymmetric Yang-Mills
theory provides a direct handle on both hydrodynamic and nonhydrodynamic
degrees of freedom in strongly coupled plasma. Understanding the dynamics of
these modes generically requires solving numerically five-dimensional Einstein
equations, which is a formidable endeavor. The goal of this Letter is to
extract the dynamics of the least damped nonhydrodynamic modes from AdS/CFT
and to incorporate them in a \emph{four-dimensional} description in which they
are coupled to conventional hydrodynamic quantities: local temperature $T$ and
flow velocity $u^\mu$.  Such a four-dimensional description should be of
definite practical utility.  Moreover, its novel structural form should have
quite general applicability.

The precise distinction between hydrodynamic and nonhydrodynamic modes is hard
to make in general, but in the hydrodynamic phase and its vicinity equilibrium
concepts are expected to approximately apply. A~natural definition of
excitations of the equilibrium plasma comes from linear response theory and is
expressed in terms of singularities of the retarded stress tensor correlator
in the complex frequency plane.

In the case of strongly coupled holographic plasma, the singularities are
single poles leading to nonequilibrium excitations characterized via complex
dispersion relations~$\omega(k)$. In a dual gravitational picture, $\omega(k)$
are the quasinormal mode (QNM) frequencies of the black brane representing
equilibrium plasma \cite{Son:2002sd,Kovtun:2005ev}. Nonhydrodynamic modes are
those which are exponentially damped for any value of momentum and, if
excited, typically become physically negligible after time of order
of~$1/\Im(\omega) $. One also finds that QNMs have $\Re(\omega) \neq 0$ and
$\Im(\omega) = {\cal O}(T)$, in line with the typical hydrodynamization scale
being ${\cal O} (1/T)$.

In this Letter, we focus on the dynamics of the mode with the smallest
non-vanishing $\Im(\omega)$, as it generically governs the direct approach to
the hydrodynamic phase. Incorporating it in a phenomenological description
including hydrodynamic modes should improve the range of applicability of such
a description. We explicitly address the case of $\nn=4$ SYM, but our
considerations should carry over verbatim to any conformal theory (or QCD in
the conformal approximation) with appropriate values of $\Re(\omega)$ and
$\Im(\omega)$ for the least damped mode.

\section{Evolution equations for quasinormal modes}

In strongly coupled field theories, expectation values of local operators,
e.g. $O = \mathrm{tr} \, F^2$ or $T^{\mu \nu}$, typically decay exponentially
when the system is perturbed out of global thermal equilibrium, with the
exception of hydrodynamic modes. The characteristic frequencies governing this
behavior can be computed as poles of the retarded Green's function and depend on
momentum. At sufficiently late times, only the lowest mode gives a physically relevant
contribution, e.g. 
\eq
\label{eq.Otx}
\langle O \rangle = \int d^{3} k A(k) e^{- \omega_{I} T t}
\cos{\left(\omega_{R} T t + \vec{k} \cdot \vec{x} + \phi(k) \right)}, 
\eqx
where $A$ and $\phi$ are some functions and we have defined
\eq
\label{eq.omega}
\omega/ T = \omega_{R}(k/T) + i \, \omega_{I}(k/T).
\eqx
In the case of holographic plasma, $\omega_{R/I}$ are given by the
quasinormal frequencies of the black brane appearing in the dual gravitational
description. Their momentum dependence has been computed 
numerically~\cite{Kovtun:2005ev} and it is apparent that both for $O$ and $T^{\mu \nu}$ they exhibit
very weak dependence on $k$ up to $k \approx 2 \pi T$. 
As far as we know, this important feature has not been emphasized so far.
This suggests
neglecting 
this dependence entirely as a first approximation, which we do in the rest of
the text.  
Under this assumption, which we will refer to as ultralocality, the
expectation value $\langle O \rangle$ given above satisfies the following
second order differential equation 
\eq
\label{eq.Oevolsimp}
\left( \frac{1}{T}
\frac{\partial}{\partial t} \right)^2  \langle O \rangle
+ 2 \, \omega_{I}
\frac{1}{T} \frac{\partial}{\partial t} \langle O \rangle +
|\omega|^2 \langle O \rangle = 0 ,
\eqx
where $|\omega|^2\equiv \omega_I^2+\omega_R^2$. 
Eq.~\rf{eq.Oevolsimp} is formally the equation of motion of a damped
harmonic oscillator. 
For ${\cal N} = 4$ SYM\footnote{Here and in the following this will always
  mean ${\cal N} = 4$ SYM theory at large $N_c$ and strong 't Hooft coupling.}
and $O = \mathrm{tr} \, F^2$ the 
frequencies (the QNM frequencies at zero momentum) are \cite{Kovtun:2005ev}
\eq
\label{eq.omegavalue}
\omega_{R} \approx 9.800 \quad \mathrm{and} \quad \omega_{I} \approx 8.629.
\eqx

The focus of interest here is the analog of eq.~\rf{eq.Oevolsimp} for the
expectation value of the energy momentum tensor, which would be a step toward
writing  phenomenological equations describing the interactions of the lowest
stress tensor QNM with the hydrodynamic degrees of freedom. To
this end note that for sufficiently near-equilibrium situations (but not
limited to hydrodynamics) the stress 
tensor can be decomposed in the following way 
\eq
\label{eq.Tmunu}
\langle T^{\mu \nu} \rangle = {\cal E} \,  u^{\mu} u^{\nu} + {\cal P}({\cal E}) (
\eta^{\mu \nu} + u^{\mu} u^{\nu} ) + \Pi^{\mu \nu} , 
\eqx
where $u_{\nu} u^{\nu} = -1$ and the symmetric tensor $\Pi^{\mu \nu}$ obeys the
Landau frame condition $u_{\mu} \Pi^{\mu \nu} = 0$. For conformal field
theories considered here one also has ${\cal P}({\cal E}) = 1/3 \, {\cal E}$, 
$\Pi^{\mu}_{\mu} = 0$. Furthermore, one defines also the ``effective temperature'' $T$ in
any state as the temperature of an equilibrium state with the same energy
density. 

In equilibrium $\Pi^{\mu \nu} = 0$ and the system can always be described in
the global rest frame, i.e. $u^{\mu} = 0$ for $\mu \neq t$ and $u^{t} =
1$. Perturbations near-equilibrium are thus $\delta T$, $\delta u^{\mu}$ with 
$\delta u^{t} = 0$ and $\delta \Pi^{\mu \nu}$ with $\delta \Pi^{t \mu} =
0$. Note that the conservation equation of the stress tensor  
\eq
\label{eq.Tmunucons}
\partial_{\mu} \langle T^{\mu \nu} \rangle = 0
\eqx 
always allow one to solve for the four variables given by $\delta T$ and $\delta
u^{\mu}$. 

At nonzero momentum, different
combinations of components of $\delta \Pi^{\mu \nu}$ (different channels)
oscillate with different frequencies. 
However, ultralocality implies that 
for momenta smaller than $k \approx 2 \pi T$ this effect is negligible and
the oscillation frequencies in all channels are approximately the same and
coincide with the frequencies in eq.~\rf{eq.omegavalue}. Because of this,
each component of $\delta \Pi^{\mu \nu}$ satisfies the same equation as
eq.~\rf{eq.Oevolsimp}:
\eq
\label{eq.Tmunuevol}
\left( \frac{1}{T}
\frac{\partial}{\partial t} \right)^2  \delta \Pi^{\mu \nu}
+ 2 \, \omega_{I}
\frac{1}{T} \frac{\partial}{\partial t} \delta \Pi^{\mu \nu} +
|\omega|^2\delta \Pi^{\mu \nu} = 0.
\eqx
Eq.~\rf{eq.Tmunuevol} together with eq.~\rf{eq.Tmunucons} describe the
evolution of the lowest nonhydrodynamic degree of freedom for small deviations
from global thermal equilibrium.

\section{Quasinormal modes in a hydrodynamic background}

In generic situations one expects that the lowest nonhydrodynamic degree of
freedom interacts with hydrodynamic modes and properly accounting for these
interactions turns out to require nontrivial modifications of
eq.~\rf{eq.Tmunuevol}. Part of these modifications can be motivated by
generalizing eq.~\rf{eq.Oevolsimp} to describe late time equilibration of
$\langle O \rangle$ on top of the plasma described by hydrodynamics, i.e. with 
\eq
\label{eq.Pimunu}
\Pi^{\mu \nu} = \Pi^{\mu \nu}_{\mathrm{hydro}} = - \eta(T) \sigma^{\mu \nu} + \ldots,
\eqx
where $\eta(T)$ is the shear viscosity, $\sigma^{\mu \nu}$ is the shear tensor
and the ellipsis denotes terms containing two and more derivatives of the
hydrodynamic 
fields. 

The naive covariantization of eq.~\rf{eq.Oevolsimp} by taking
$\partial_{t} \rightarrow u^{\mu} \partial_{\mu}$ and using $T$ and $u^{\mu}$
solving eq.~\rf{eq.Tmunucons} with $\Pi^{\mu \nu}$ in hydrodynamic form,
does not preserve the Weyl-covariance of the microscopic
theory.\footnote{We neglect here the effects of the Weyl-anomaly, as in
  \cite{Friess:2006kw,Baier:2007ix}.} The latter is the 
statement that under Weyl-rescaling of the background metric 
\eq
\eta_{\mu \nu} \rightarrow e^{-2\omega(x)} \eta_{\mu \nu}
\eqx
both $O$ and $T^{\mu \nu}$ transform homogeneously. In general, a field $\phi$
is said to transform with Weyl weight $w$ if
\eq
\phi \rightarrow e^{w \omega(x)} \phi .
\eqx
Thus, for example, the metric components $g_{\mu\nu}$ transform with weight
$-2$, while $T^{\mu \nu}$ transform with weight $6$. 

These properties have led to the development of the Weyl-covariant 
formulation \cite{Loganayagam:2008is}, in which the equations of conformal
hydrodynamics assume a very compact form. This formalism makes use of the
(nondynamical) ``Weyl connection''
\eq
{\cal A}_{\mu} = u^{\lambda} \nabla_{\lambda} u_{\mu} - \frac{1}{3} \nabla_{\lambda} u^{\lambda} u_{\mu}.
\eqx
to define a derivative operator, denoted here by
$\D_\mu$, which is covariant under Weyl-transformations\footnote{A general
  formula can be found in \cite{Loganayagam:2008is}.}. 

We have checked, by performing an explicit gravitational calculation of the
lowest quasinormal mode in the viscous fluid background
\cite{Bhattacharyya:2008jc}, that the covariantization of eq.~\rf{eq.Oevolsimp}
with the use of the Weyl-covariant derivative, i.e. $\partial_{t} \rightarrow
\D\equiv u^{\mu} \D_{\mu}$, reproduces the correct result. Hence, the natural
generalization 
of eq.~\rf{eq.Tmunuevol} is 
\eq
\label{eq.TmunuevolWeyl}
(\f{1}{T} \D)^2  \qnm_{\mu\nu} +2\omega_I \f{1}{T} \D \qnm_{\mu\nu}+
|\omega|^2 \qnm_{\mu\nu}=0,
\eqx
where the role of $\delta\Pi$ is now taken on by 
\eq\label{eq.deltapi}
\qnm^{\mu \nu} = \Pi^{\mu \nu} - \Pi_{\mathrm{hydro}}^{\mu \nu} 
\eqx
and 
\eqn
\label{eq.Weyl-covderiv}
\D  \qnm_{\mu\nu} &=&  u^\lambda \left(\nabla_\lambda + 4 {\cal
  A}_{\lambda}\right)  \qnm_{\mu\nu} -  2{\cal A}_{\lambda} u^{(\mu}
  \qnm^{\nu)\lambda}
\eqnx
This formula also defines the action of $\D$ on $\f{1}{T}\D \qnm^{\mu \nu}$,
since the latter object has the same Weyl weight as  $\qnm_{\mu\nu}$.

Equation \rf{eq.TmunuevolWeyl} has two key features. First, it is consistent with
$\Pi^{\mu  \nu}$ transforming homogeneously under Weyl
transformations. Secondly, it preserves its transversality and tracelessness
due to the fact that $\D \, u^\mu = 0$.

As a nontrivial test of equation~\rf{eq.TmunuevolWeyl} we have checked that
it is obeyed by the QNM computed in~\cite{Heller:2013fn} for the strongly
coupled plasma undergoing Bjorken expansion \cite{Bjorken:1982qr}. Even though this is a special
flow with a high degree of symmetry, already in this case the terms coming
from the Weyl connection are nontrivial.

\section{Generalized theories of hydrodynamics}

The Landau-Lifschitz theory of relativistic viscous hydrodynamics is defined
by adopting as the evolution equation the conservation of the stress
tensor~\rf{eq.Tmunu} with $\Pi^{\mu \nu}$ given by
eq.~\rf{eq.Pimunu}. However this system of differential equations is not
hyperbolic and in general does not have a well-posed initial value
problem~\cite{Hiscock:1985zz,PhysRevD.62.023003}.

Hyperbolic theories of hydrodynamics, postulated by M{\"u}ller, Israel,
Stewart and others \cite{Muller:1967zza,Israel:1979wp}, instead of
using \rf{eq.Pimunu} assume that the shear tensor is replaced by a new dynamical
object, 
$\Pi_{MIS}^{\mu \nu}$ which obeys an evolution equation
involving 
additional phenomenological parameters. A prototypical example of such an
equation is 
\eq
\label{eqMIS}
\left( \hat{\tau}_{\Pi} \f{1}{T} \D + 1 \right)
\Pi_{MIS}^{\mu \nu} = - \eta \sigma^{\mu\nu} \ ,
\eqx
where $\hat{\tau}_{\Pi}$ is a dimensionless constant and the combination
$\hat{\tau}_{\Pi} \f{1}{T}$ has been referred to in the 
literature as the relaxation time. Eq.~\rf{eqMIS} can be supplemented with
terms quadratic in $\Pi^{\mu \nu}$ and gradients of hydrodynamic fields in
such a way that solving it recursively in the gradient expansion gives the
correct form of the hydrodynamic stress tensor up to second order in
derivatives~\cite{Baier:2007ix} (when referring to MIS theory in the following
we will always mean this formulation). 
In this approach the relaxation time is
identified with one of the second order transport coefficients.
Assuming $\eta/s=1/(4\pi)$, the linearized theory is causal as long as
$\hat{\tau}_{\Pi}\geq 1/(2\pi)$. The drawback of the MIS formulation, however, is
that 
it introduces a spurious nonphysical decaying mode with a frequency given by
the relaxation time: $\omega = i \, T / \hat{\tau}_{\Pi}$.

The simplest way to incorporate additional \emph{physical} nonequilibrium
degrees of freedom into a causal hyperbolic description is to set
\eq
\label{newtmunu}
\Pi^{\mu \nu} =  \Pi_{MIS}^{\mu \nu} + \qnm^{\mu \nu}
\eqx
with $\Pi_{MIS}^{\mu \nu}$ satisfying~\rf{eqMIS} and $\qnm^{\mu \nu}$ 
obeying~\rf{eq.TmunuevolWeyl}. These traceless and transverse quantities are coupled together
by the conservation law~\rf{eq.Tmunucons}.

The resulting theory satisfies the same causality and stability properties as
the MIS formulation.  At the linearized level, in addition to the standard
hydrodynamic modes it contains the damped modes corresponding to QNM as seen
in AdS/CFT. However, as a byproduct of using the MIS formulation we have in
addition the spurious decaying mode of MIS theory discussed above.  In order
to minimize its impact, we always set $-\eta(T) \sigma^{\mu\nu}$ as the
initial condition for $\Pi_{MIS}^{\mu \nu}$.  Moreover we set the $\tau_\Pi$
parameter to the smallest value allowed by causality in order to maximize the
damping of this mode.

The above formulation is the simplest generalization of MIS hydrodynamics.
The equations presented here should provide a useful extension of
hydrodynamics in situations, where only a single QNM dominates the approach to
equilibrium.  Setting vanishing initial conditions for $\qnm^{\mu \nu}$
reduces the theory to standard MIS, while incorporating some nontrivial
initial conditions allows us to examine the physical effects of the least
damped nonhydrodynamic degrees of freedom.  This theory could be used as an
alternative to MIS hydrodynamics in situations, when an account of early
pre-equilibrium dynamics including modes with $\Re{(\omega)} \neq 0$ is
relevant.  We perform various tests of this theory in the following section.

Before that however, we would like to mention a possible alternative
which aims to get rid of the nonphysical MIS mode altogether and
use the physical nonequilibrium degrees of freedom as a means of ensuring
hyperbolicity.  Note that since the QNM have a sizable real
frequency, one can never describe them using the MIS decaying mode. This has
already been emphasized in~\cite{Noronha:2011fi}.

Heuristically one could proceed by
using eq.~\rf{eq.deltapi} and~\rf{eq.Pimunu} in eq.~\rf{eq.TmunuevolWeyl} to
find  
\eqn
\label{eqpi2s}
 \left((\f{1}{T} \D)^2\right. &+& \left. 2\omega_I \f{1}{T} \D + |\omega|^2\right)
\Pi^{\mu \nu} =  \nonumber\\
&-& \eta  |\omega|^2 \sigma^{\mu\nu}  - c_\sigma \f{1}{T} \D\left(\eta
\sigma^{\mu\nu}\right) + \ldots 
\eqnx
where the ellipsis denotes contributions of second and higher order in
gradients. Of all possible second order terms only one term has been kept,
with a  
coefficient $c_\sigma$, which is treated as an 
arbitrary parameter\footnote{Solving eq.~\rf{eq.Pimunu} in the gradient
  expansion shows that $c_\sigma$ contributes to second order transport
  coefficients.}. 
This term is included explicitly, since it improves 
the stability of~\rf{eqpi2s}.

The key property of eq.~\rf{eqpi2s} is that linearization around an
equilibrium background leads to a system of partial differential equations
which is hyperbolic for $c_\sigma\geq 0$. 
The characteristic velocity in the
sound channel is found to be
\eq
v = \frac{1}{\sqrt{3}} \left(1 + \frac{c_{\sigma}}{\pi} \right)^{1/2}
, 
\eqx
so for causality one must further impose $c_\sigma \leq 2\pi$ (this in
fact ensures causality in all channels). 

For a numerical treatment of  Eq.~\rf{eqpi2s} it is important that
exponentially growing modes be absent. Whether Eq.~\rf{eqpi2s} is stable in
this sense depends on the values of parameters such as the QNM frequencies and 
the viscosity to entropy ratio. This is similar the case the MIS
equations. However, unlike that case, for the values of $\eta/s$ and
$\omega_{R,I}$ characteristic of $\nn=4$ SYM, eq.~\rf{eqpi2s} contains
exponentially unstable modes with high $k$. This renders these equations 
(as they stand) unsuitable for numerical evaluation and comparison to the
results of simulations based on the AdS/CFT correspondence. Let us emphasize,
however, that these unstable modes appear far outside the range of
applicability of the long wavelength description (e.g. with wave vectors $k>18.5
T$ if one chooses $c_\sigma =2\pi$). 
It would be
interesting to investigate whether one could modify Eq.~\rf{eqpi2s} to cure this
pathology. This question is set aside for the moment, and we henceforth
concentrate on the simplest formulation given by Eq.~\rf{newtmunu} and
Eq.~\rf{eq.TmunuevolWeyl}.

\section{Tests}

An essential part of this Letter is testing 
the equations \rf{newtmunu} and 
\rf{eq.TmunuevolWeyl}, \rf{eqMIS} against microscopic
numerical computations of ${\cal N} = 4$ 
SYM plasma based on the AdS/CFT
correspondence. This requires setting the parameters to 
appropriate values, i.e. $\eta/s = 1/4\pi$ and
$\omega_{R,I}$ as in eq.~\rf{eq.omegavalue}. 
We also set $\tau_\Pi=1/(2\pi)$, which is the smallest
value allowed by causality.

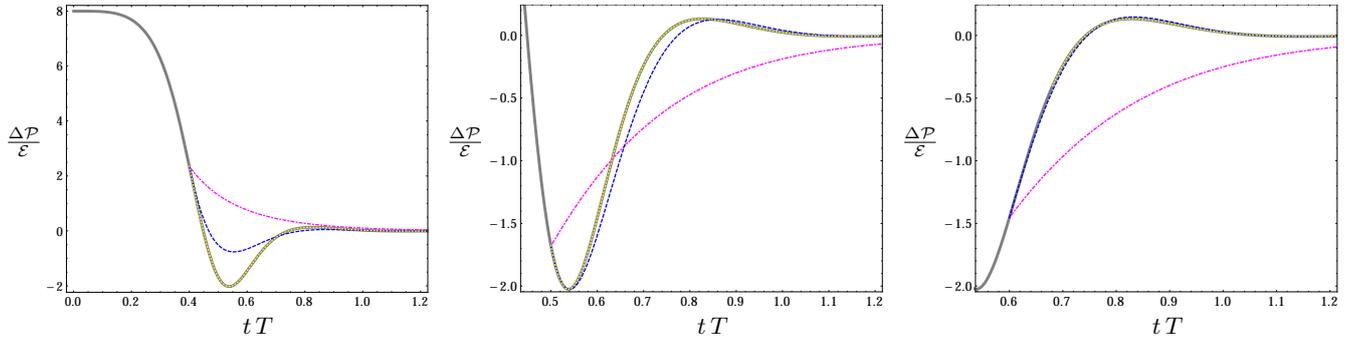
\begin{figure*}[t]
\begin{tabular}{ccc}
\includegraphics[height=0.17\textheight]{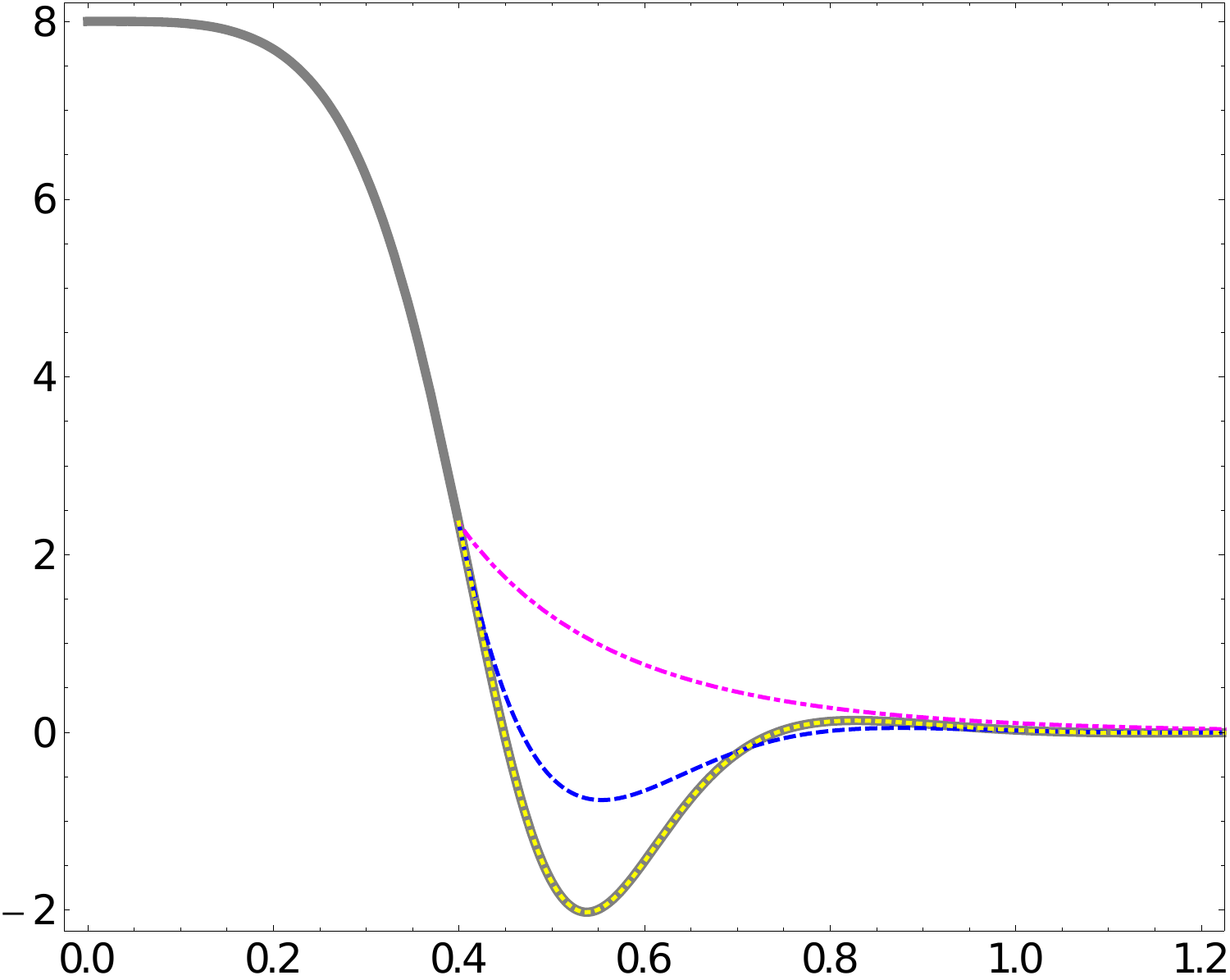}
\put(-159,60){$\frac{\Delta {\cal P}}{{\cal E}}$}
\put(-70,-10){$t \, T$}
\quad & \quad \quad
\includegraphics[height=0.17\textheight]{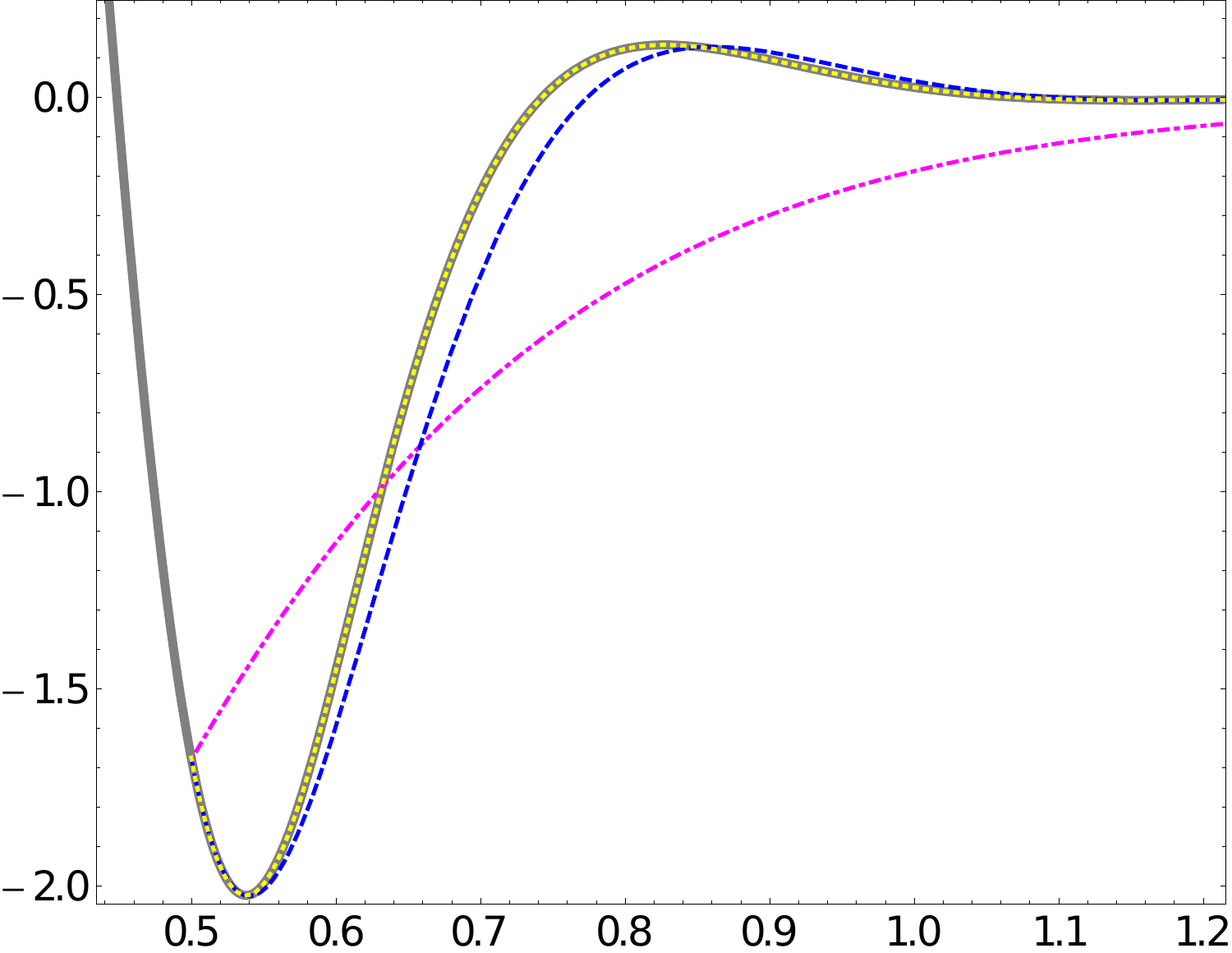}
\put(-164,60){$\frac{\Delta {\cal P}}{{\cal E}}$}
\put(-70,-10){$t \, T$}
\quad & \quad \quad
\includegraphics[height=0.17\textheight]{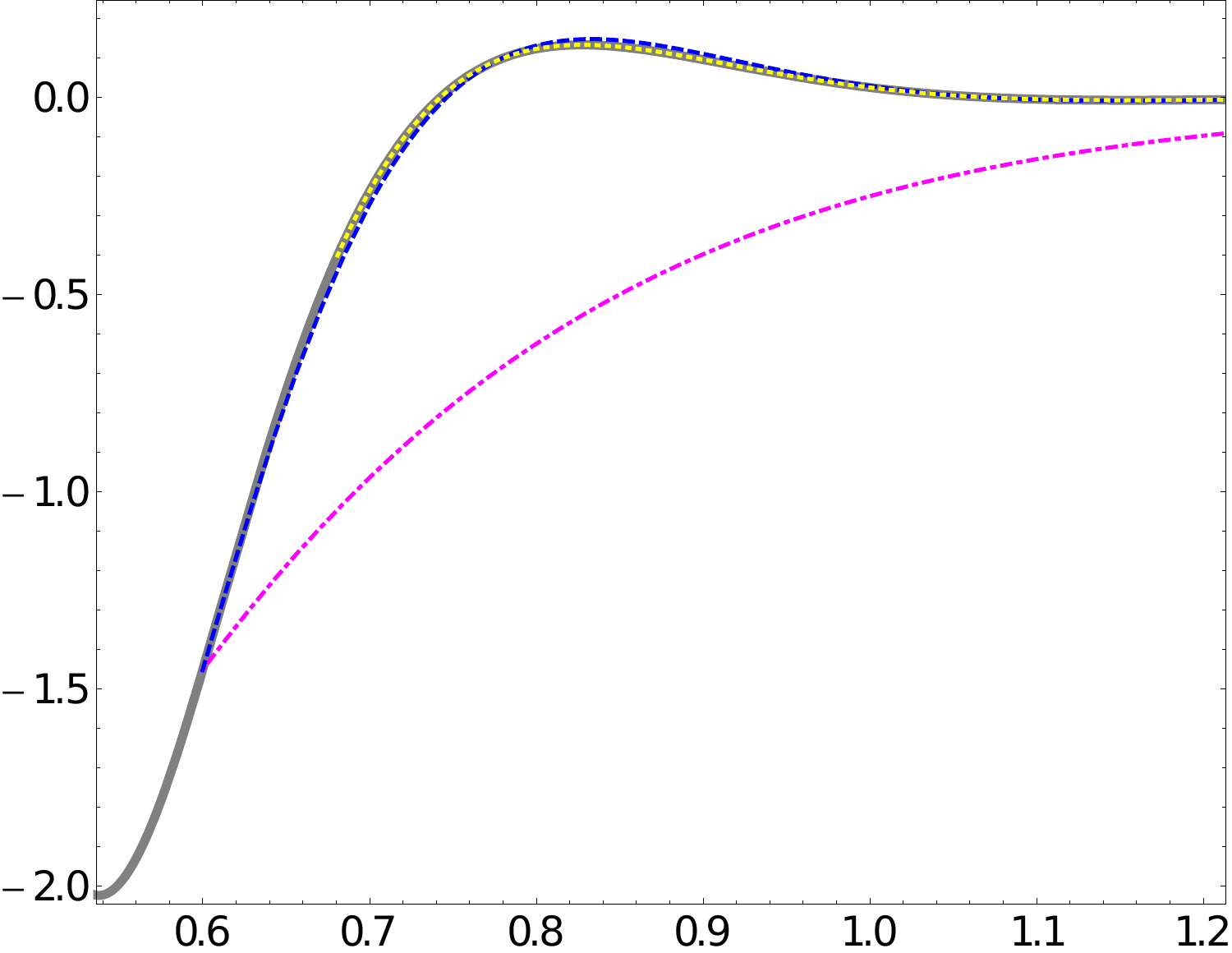}
\put(-164,60){$\frac{\Delta {\cal P}}{{\cal E}}$}
\put(-70,-10){$t \, T$}
\end{tabular}
    \hfill
\caption{
The stress tensor for homogeneous isotropization is $\langle T^{\mu \nu} \rangle  = \mathrm{diag} \left( {\cal E},
  \, \frac{1}{3} {\cal E} - \frac{2}{3} \Delta {\cal P}(t), \,  \frac{1}{3}
     {\cal E} + \frac{1}{3} \Delta {\cal P}(t), \,  \frac{1}{3} {\cal E} +
     \frac{1}{3} \Delta {\cal P}(t) \right)^{\mu \nu}$. The effective
     temperature remains constant throughout evolution, which is entirely
     specified by providing $\cal E$ or $T$ and $\Delta {\cal P}(t)$.
 The gray
     curve denotes the numerical solution of Einstein's equations. 
     The dotted yellow curves denote solutions of
     linearized Einstein's equations (including all the QNM),
     dot-dashed magenta curves are solutions of MIS theory and the 
     dashed blue curves represent solutions of the new theory given by
     eq.~\rf{newtmunu}. The 
     initialization times are $t T = 0.4$ (left), $0.5$ (centre) and $0.6$ (right). 
     The sum of the QNM
     describes the 
     evolution of the system very well already at $t T = 0.3$. This is in line
     with the findings in \cite{Heller:2012km,Heller:2013oxa}. For late enough
     initialization 
the new equations do a much better job
     in describing the dynamics of the pressure anisotropy than MIS theory,
     which can underestimate the isotropization time by more than a
     factor of $2$ (centre). } 
\label{figiso}
\end{figure*}

Here we consider two particularly symmetric configurations: homogeneous
isotropization and boost-invariant flow.  It is worth emphasizing at this
point that homogeneous isotropization cannot be described at all by
conventional Landau-Lifshitz viscous hydrodynamics.

The AdS/CFT computations are based on numerical solutions of
$(4+1)$-dimensional Einstein's equations with negative cosmological constant
obtained following the methods developed in
\cite{Heller:2012km,Heller:2013oxa} and
\cite{Heller:2011ju,HellerSpalinski}. This we compare to numerical solutions
of the new phenomenological equations initialized by
specifying just the energy, pressure anisotropy and its time derivative which
we take to agree with the values extracted from a particular numerical
solution of Einstein equations at the specific initialization time.

The results for holographic isotropization, depicted on Fig.~\ref{figiso},
show that for late enough initialization, eq.~\rf{newtmunu} captures both the
qualitative and quantitative features of the pressure anisotropy
relaxation. Comparison to a solution of linearized Einstein's equations, which
can be superficially thought of as a sum over all 
quasinormal modes in this system, demonstrates that the applicability of the
new equations is not limited by the far-from-equilibrium nonlinear effects not
captured by it, but rather by the presence of the higher quasinormal modes (as
clearly seen in the center and right plots in Fig.~\ref{figiso}).

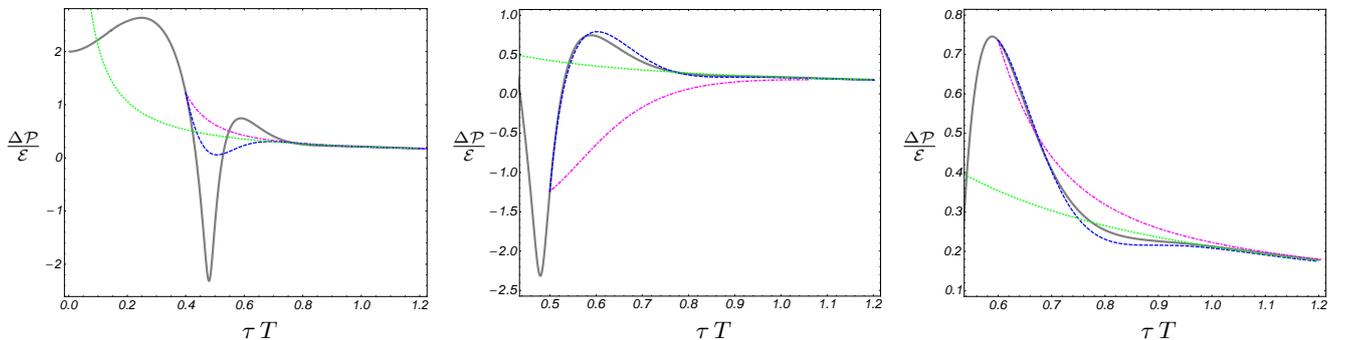
\begin{figure*}[t]
\begin{tabular}{ccc}
\includegraphics[height=0.17\textheight]{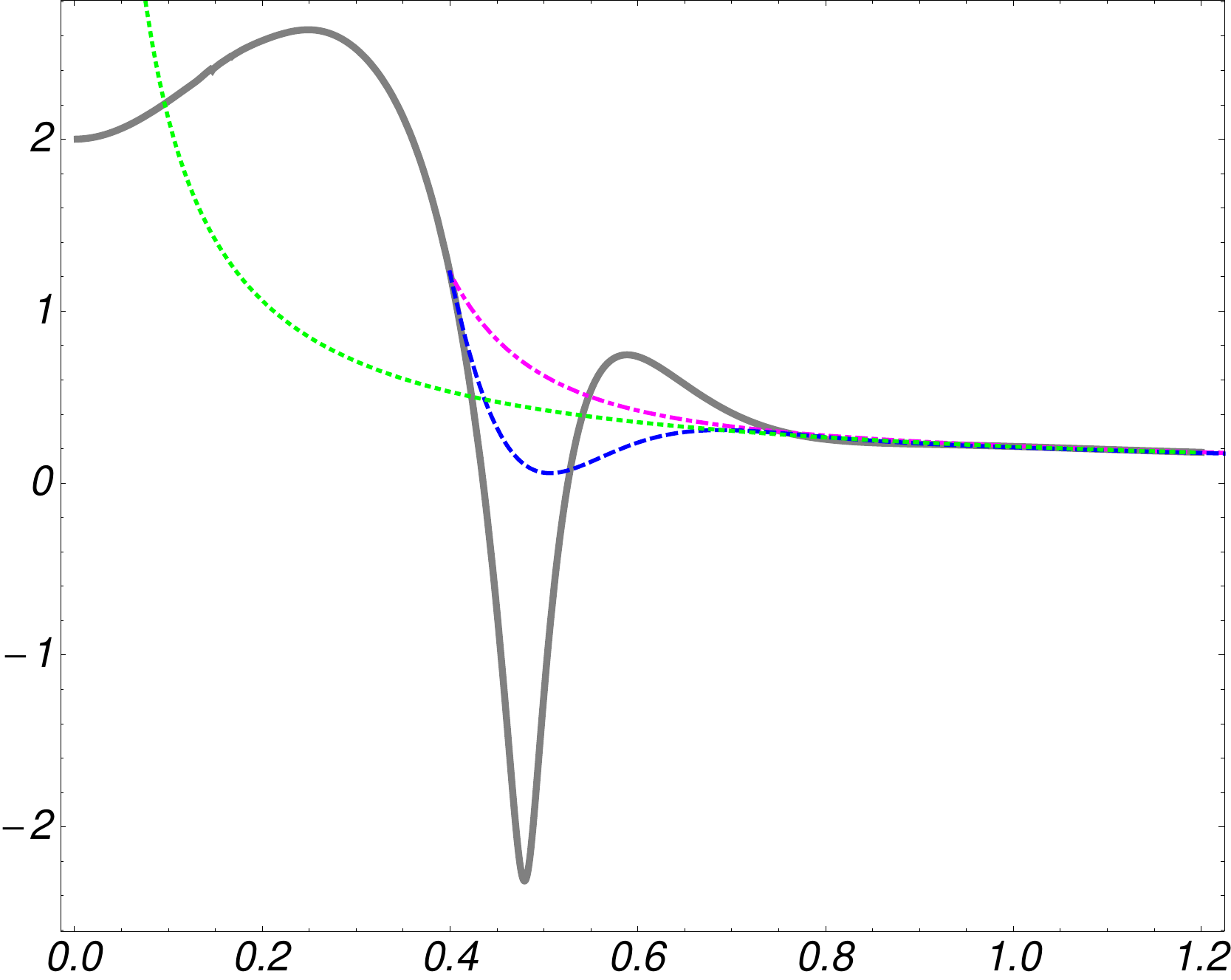}
\put(-159,60){$\frac{\Delta {\cal P}}{{\cal E}}$}
\put(-70,-10){$\tau \, T$}
\quad & \quad \quad
\includegraphics[height=0.17\textheight]{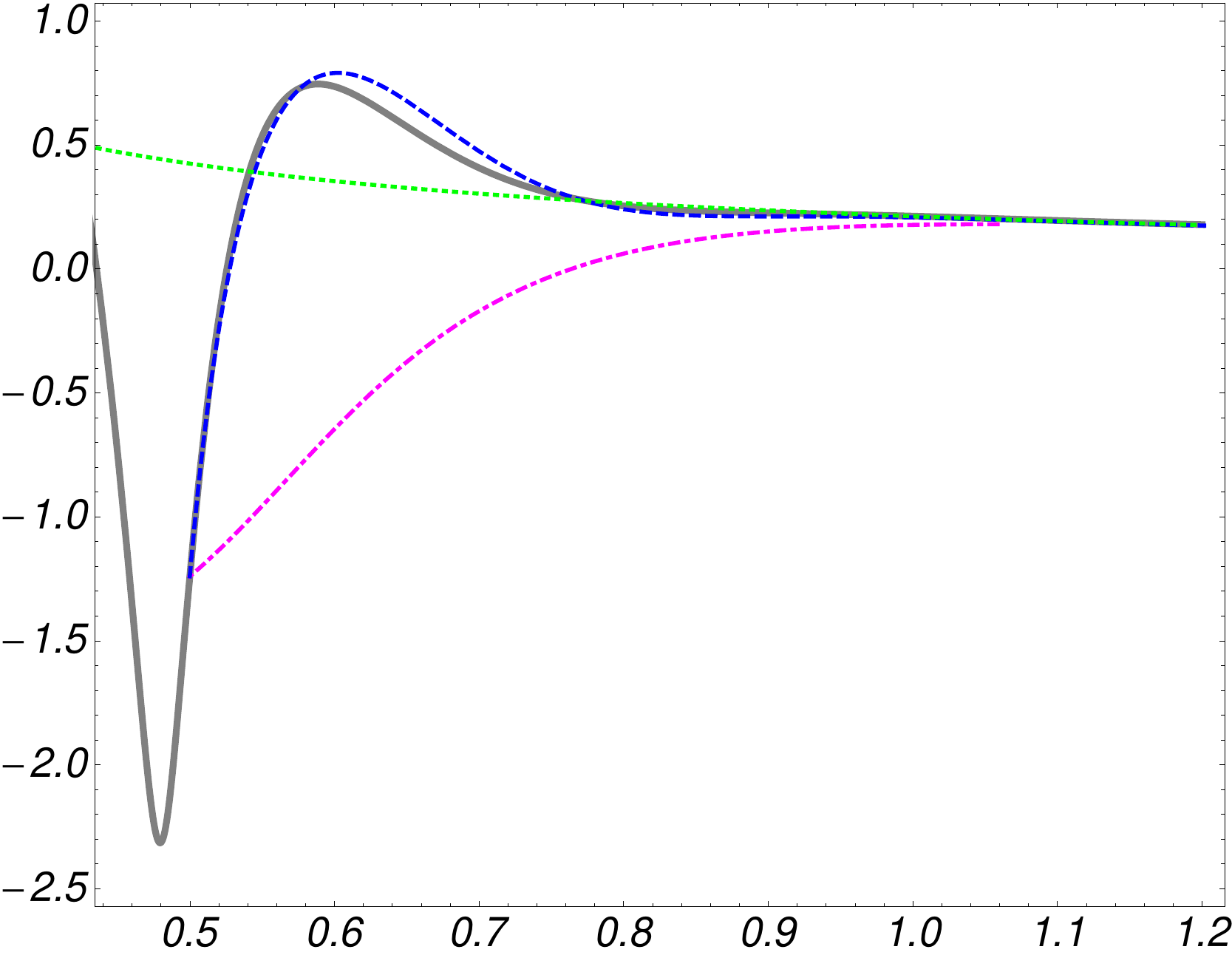}
\put(-162,60){$\frac{\Delta {\cal P}}{{\cal E}}$}
\put(-70,-10){$\tau \, T$}
\quad & \quad \quad
\includegraphics[height=0.17\textheight]{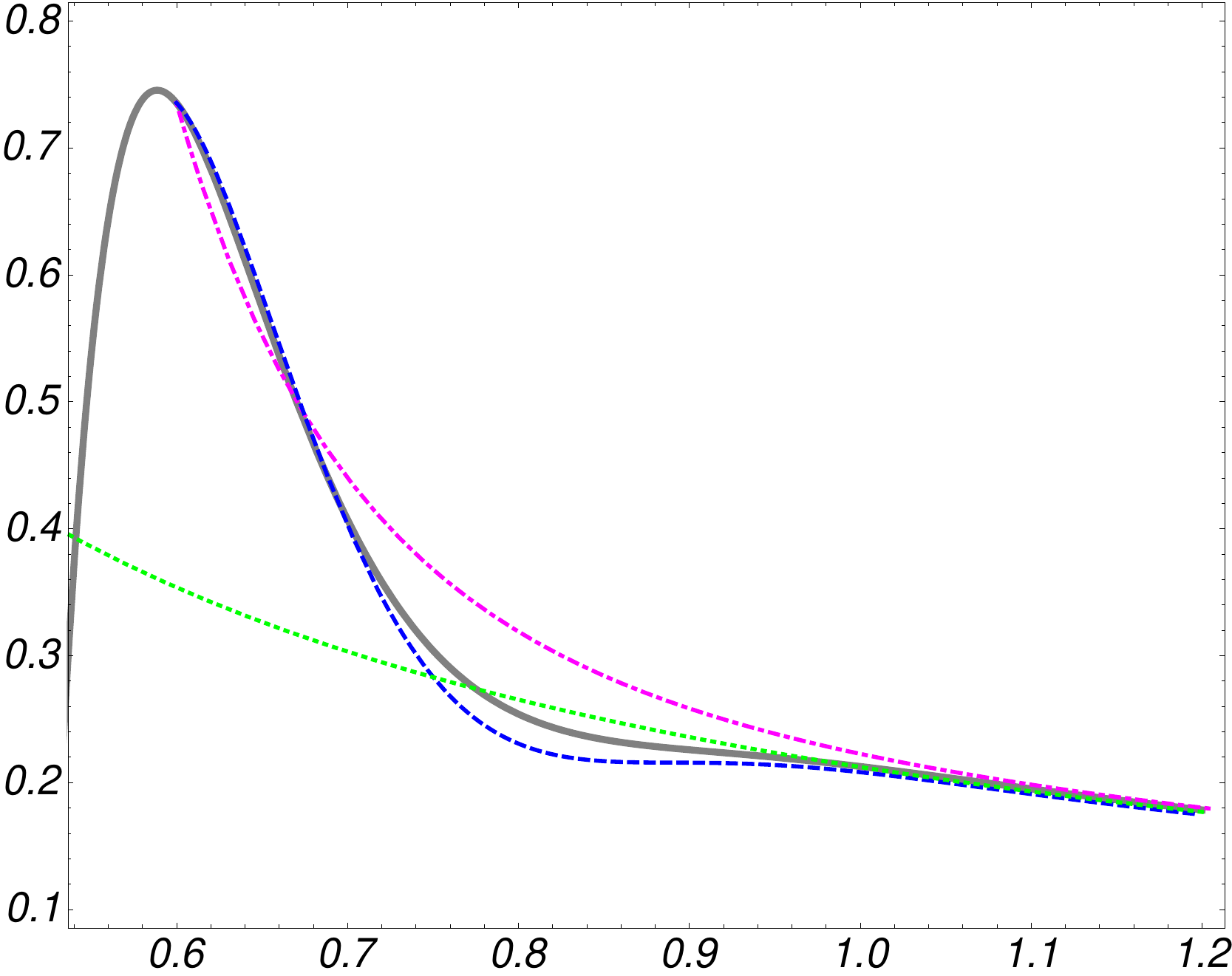}
\put(-160,60){$\frac{\Delta {\cal P}}{{\cal E}}$}
\put(-70,-10){$\tau \, T$}
\end{tabular}
    \hfill
\caption{Boost-invariant flow is a one-dimensional expansion of plasma in
  which the late time behavior is dominated by a hydrodynamic tail. In the
  local rest frame (proper time $\tau$ -- rapidity $y$ coordinates) the 
  stress tensor takes the form $\langle T^{\tau \tau} \rangle =
  {\cal E}(\tau)$, $\langle T^{y y} \rangle = \tau^{-2} {\cal P}_{L}(\tau)$ and 
  $\langle T^{\perp \perp} \rangle = {\cal P}_{T}(\tau)$. The plots depict the pressure
  anisotropy $\Delta {\cal P} \equiv ({\cal P}_T- {\cal P}_L)$ normalized by ${\cal E}$.
Gray curves denote the numerical solution based on AdS/CFT; magenta
dash-dotted curves are solutions of MIS theory and the blue dashed curves 
  are solutions of the new theory defined via eq.~\rf{newtmunu}. For
  reference, the prediction of first order hydrodynamics is displayed as the
  dotted green curve. The plots show the results of setting initial conditions at 
$\tau \, T = 0.4$ (left), $0.5$ (centre) and $0.6$
  (right). One 
  can see that both MIS theory and the new equations converge to the exact
  curve at late times, which demonstrates the applicability of
  viscous hydrodynamics. With earlier initialization (center), the new 
  equations lead to a quantitative agreement with the data also in the
  pre-equilibrium phase, as opposed to the MIS description.} 
\label{figbif}
\end{figure*} 

The case of boost-invariant flow is presented in Fig.~\ref{figbif}, which
shows clearly that the MIS approach captures the late time tail very well, as
do the new equations proposed here. However, at earlier times eq.~\rf{newtmunu}
provides a much more accurate picture. Estimates
of the final temperature are also more accurate if eq.~\rf{newtmunu} is
used. For initial conditions involving many
QNMs the agreement at early times should not be as good (in analogy with what
is seen in Fig.~\ref{figiso}). Also, for initial 
conditions where no nohydrodynamic modes are excited at early times, 
effects of second and higher order (or possibly resummed
\cite{Lublinsky:2007mm}) hydrodynamics may become important.

\section{Summary and conclusions}

The new phenomenological equations presented in this Letter generalize
relativistic Navier Stokes theory by including leading nonhydrodynamic modes
expected in theories of strongly coupled plasma with gravity duals. In these
theories the nonhydrodynamic modes correspond to QNMs of
black branes in asymptotically AdS space. The weak dependence of QNM
frequencies on momenta suggests the ultralocality assumption,
which we have used to identify the second order equation satisfied by the QNM
contribution to the shear stress tensor. This equation is the essential new
element, which makes it possible to go beyond the observations made in
reference 
\cite{Denicol:2011fa,Noronha:2011fi}, where generalizations of
hydrodynamics were 
pursued having
noted the significance of the analytic structure of retarded correlators in
theories with gravity duals.

The use of a conventional hydrodynamic description implicitly assumes that all
nonequilibrium collective excitations in the quark-gluon plasma are 
set to zero. The proposed equations provide a means of relaxing this
assumption and exploring their influence on subsequent hydrodynamic evolution.
For some observables (such as the final multiplicities) this may not be
quantitatively important.  However for observables sensitive to the
pre-equilibrium stages of evolution (such as photon
\cite{Schenke:2006yp,McLerran:2014hza} or dilepton emission
\cite{Martinez:2008di,Shuryak:2012nf}) capturing the early time dynamics may
be valuable. 
An important step toward such applications will be to develop an effective
heuristic for setting initial conditions for the nonhydrodynamic modes in our
new evolution equations. One of the possible approaches might be to extract
these initial conditions from the early post-collision state following from
the numerical simulations of \cite{Gelis:2013rba} or \cite{Berges:2013fga}.

\smallskip

\noindent{\bf Acknowledgments.}  
We would like to 
thank U.~Heinz, M.~Martinez, S.~Mr\'owczynski, J.~Noronha, R.~Peschanski,
K.~Rajagopal, P.~Romatschke, W.~van der Schee, M.~Strickland, D.~Teaney,
A.~Vuorinen, U.~Wiedemann, L.~Yaffe and A.~Yarom for valuable 
discussions and correspondence. MPH is supported by the Netherlands
Organization for Scientific Research under the NWO Veni scheme (UvA). This
work was supported by the 
National Science Centre grants 2012/07/B/ST2/03794
(MPH and MS), 2012/06/A/ST2/00396 (RJ) and 2013/11/N/ST2/03812 (PW). Research
at Perimeter Institute is supported by the Government of Canada through
Industry Canada and by the Province of Ontario through the Ministry of
Research \& Innovation.

\bibliography{genmis_biblio}{}

\providecommand{\href}[2]{#2}\begingroup\raggedright\begin{thebibliography}{10}

\bibitem{Heinz:2004pj}
U.~W. Heinz, ``{Thermalization at RHIC},''
  \href{http://dx.doi.org/10.1063/1.1843595}{{\em AIP Conf.Proc.} {\bf 739}
  (2005)  163--180},
\href{http://arxiv.org/abs/nucl-th/0407067}{{\tt arXiv:nucl-th/0407067
  [nucl-th]}}.

\bibitem{Wiedemann:2012py}
U.~A. Wiedemann, ``{Introductory Overview of Quark Matter 2012},''
  \href{http://dx.doi.org/10.1016/j.nuclphysa.2013.01.038}{{\em Nucl.Phys.}
  {\bf A904-905} (2013)  3c--10c},
\href{http://arxiv.org/abs/1212.3306}{{\tt arXiv:1212.3306 [hep-ph]}}.

\bibitem{Chesler:2009cy}
P.~M. Chesler and L.~G. Yaffe, ``{Boost invariant flow, black hole formation,
  and far-from-equilibrium dynamics in N = 4 supersymmetric Yang-Mills
  theory},'' \href{http://dx.doi.org/10.1103/PhysRevD.82.026006}{{\em
  Phys.Rev.} {\bf D82} (2010)  026006},
\href{http://arxiv.org/abs/0906.4426}{{\tt arXiv:0906.4426 [hep-th]}}.

\bibitem{Chesler:2010bi}
P.~M. Chesler and L.~G. Yaffe, ``{Holography and colliding gravitational shock
  waves in asymptotically AdS$_5$ spacetime},''
  \href{http://dx.doi.org/10.1103/PhysRevLett.106.021601}{{\em Phys.Rev.Lett.}
  {\bf 106} (2011)  021601},
\href{http://arxiv.org/abs/1011.3562}{{\tt arXiv:1011.3562 [hep-th]}}.

\bibitem{Heller:2011ju}
M.~P. Heller, R.~A. Janik, and P.~Witaszczyk, ``{The characteristics of
  thermalization of boost-invariant plasma from holography},''
  \href{http://dx.doi.org/10.1103/PhysRevLett.108.201602}{{\em Phys.Rev.Lett.}
  {\bf 108} (2012)  201602},
\href{http://arxiv.org/abs/1103.3452}{{\tt arXiv:1103.3452 [hep-th]}}.

\bibitem{Casalderrey-Solana:2013aba}
J.~Casalderrey-Solana, M.~P. Heller, D.~Mateos, and W.~van~der Schee, ``{From
  full stopping to transparency in a holographic model of heavy ion
  collisions},'' \href{http://dx.doi.org/10.1103/PhysRevLett.111.181601}{{\em
  Phys.Rev.Lett.} {\bf 111} (2013)  181601},
\href{http://arxiv.org/abs/1305.4919}{{\tt arXiv:1305.4919 [hep-th]}}.

\bibitem{Son:2002sd}
D.~T. Son and A.~O. Starinets, ``{Minkowski space correlators in AdS / CFT
  correspondence: Recipe and applications},''
  \href{http://dx.doi.org/10.1088/1126-6708/2002/09/042}{{\em JHEP} {\bf 0209}
  (2002)  042},
\href{http://arxiv.org/abs/hep-th/0205051}{{\tt arXiv:hep-th/0205051
  [hep-th]}}.

\bibitem{Kovtun:2005ev}
P.~K. Kovtun and A.~O. Starinets, ``{Quasinormal modes and holography},''
  \href{http://dx.doi.org/10.1103/PhysRevD.72.086009}{{\em Phys.Rev.} {\bf D72}
  (2005)  086009},
\href{http://arxiv.org/abs/hep-th/0506184}{{\tt arXiv:hep-th/0506184
  [hep-th]}}.

\bibitem{Friess:2006kw}
J.~J. Friess, S.~S. Gubser, G.~Michalogiorgakis, and S.~S. Pufu, ``{Expanding
  plasmas and quasinormal modes of anti-de Sitter black holes},''
  \href{http://dx.doi.org/10.1088/1126-6708/2007/04/080}{{\em JHEP} {\bf 0704}
  (2007)  080},
\href{http://arxiv.org/abs/hep-th/0611005}{{\tt arXiv:hep-th/0611005
  [hep-th]}}.

\bibitem{Baier:2007ix}
R.~Baier, P.~Romatschke, D.~T. Son, A.~O. Starinets, and M.~A. Stephanov,
  ``{Relativistic viscous hydrodynamics, conformal invariance, and
  holography},'' \href{http://dx.doi.org/10.1088/1126-6708/2008/04/100}{{\em
  JHEP} {\bf 0804} (2008)  100},
\href{http://arxiv.org/abs/0712.2451}{{\tt arXiv:0712.2451 [hep-th]}}.

\bibitem{Loganayagam:2008is}
R.~Loganayagam, ``{Entropy Current in Conformal Hydrodynamics},''
  \href{http://dx.doi.org/10.1088/1126-6708/2008/05/087}{{\em JHEP} {\bf 0805}
  (2008)  087},
\href{http://arxiv.org/abs/0801.3701}{{\tt arXiv:0801.3701 [hep-th]}}.

\bibitem{Bhattacharyya:2008jc}
S.~Bhattacharyya, V.~E. Hubeny, S.~Minwalla, and M.~Rangamani, ``{Nonlinear
  Fluid Dynamics from Gravity},''
  \href{http://dx.doi.org/10.1088/1126-6708/2008/02/045}{{\em JHEP} {\bf 0802}
  (2008)  045},
\href{http://arxiv.org/abs/0712.2456}{{\tt arXiv:0712.2456 [hep-th]}}.

\bibitem{Heller:2013fn}
M.~P. Heller, R.~A. Janik, and P.~Witaszczyk, ``{Hydrodynamic Gradient
  Expansion in Gauge Theory Plasmas},''
  \href{http://dx.doi.org/10.1103/PhysRevLett.110.211602}{{\em Phys.Rev.Lett.}
  {\bf 110} (2013) no.~21, 211602},
\href{http://arxiv.org/abs/1302.0697}{{\tt arXiv:1302.0697 [hep-th]}}.

\bibitem{Bjorken:1982qr}
J.~Bjorken, ``{Highly Relativistic Nucleus-Nucleus Collisions: The Central
  Rapidity Region},''
\href{http://dx.doi.org/10.1103/PhysRevD.27.140}{{\em Phys.Rev.} {\bf D27}
  (1983)  140--151}.

\bibitem{Hiscock:1985zz}
W.~A. Hiscock and L.~Lindblom, ``{Generic instabilities in first-order
  dissipative relativistic fluid theories},''
\href{http://dx.doi.org/10.1103/PhysRevD.31.725}{{\em Phys.Rev.} {\bf D31}
  (1985)  725--733}.

\bibitem{PhysRevD.62.023003}
P.~Kost{\"a}dt and M.~Liu, ``{Causality and stability of the relativistic
  diffusion equation},''
  \href{http://dx.doi.org/10.1103/PhysRevD.62.023003}{{\em Phys.Rev.} {\bf D62}
  (2000)  023003}, \href{http://arxiv.org/abs/cond-mat/0010276}{{\tt
  arXiv:cond-mat/0010276}}.

\bibitem{Muller:1967zza}
I.~Muller, ``{Zum Paradoxon der Warmeleitungstheorie},''
\href{http://dx.doi.org/10.1007/BF01326412}{{\em Z.Phys.} {\bf 198} (1967)
  329--344}.

\bibitem{Israel:1979wp}
W.~Israel and J.~Stewart, ``{Transient relativistic thermodynamics and kinetic
  theory},''
\href{http://dx.doi.org/10.1016/0003-4916(79)90130-1}{{\em Annals Phys.} {\bf
  118} (1979)  341--372}.

\bibitem{Noronha:2011fi}
J.~Noronha and G.~S. Denicol, ``{Transient Fluid Dynamics of the Quark-Gluon
  Plasma According to AdS/CFT},''
\href{http://arxiv.org/abs/1104.2415}{{\tt arXiv:1104.2415 [hep-th]}}.

\bibitem{Heller:2012km}
M.~P. Heller, D.~Mateos, W.~van~der Schee, and D.~Trancanelli, ``{Strong
  Coupling Isotropization of Non-Abelian Plasmas Simplified},''
  \href{http://dx.doi.org/10.1103/PhysRevLett.108.191601}{{\em Phys.Rev.Lett.}
  {\bf 108} (2012)  191601},
\href{http://arxiv.org/abs/1202.0981}{{\tt arXiv:1202.0981 [hep-th]}}.

\bibitem{Heller:2013oxa}
M.~P. Heller, D.~Mateos, W.~van~der Schee, and M.~Triana, ``{Holographic
  isotropization linearized},''
  \href{http://dx.doi.org/10.1007/JHEP09(2013)026}{{\em JHEP} {\bf 1309} (2013)
   026},
\href{http://arxiv.org/abs/1304.5172}{{\tt arXiv:1304.5172 [hep-th]}}.

\bibitem{HellerSpalinski}
M.~P. Heller and M.~Spali{\'n}ski, ``{Unpublished},''.

\bibitem{Lublinsky:2007mm}
M.~Lublinsky and E.~Shuryak, ``{How much entropy is produced in strongly
  coupled Quark-Gluon Plasma (sQGP) by dissipative effects?},''
  \href{http://dx.doi.org/10.1103/PhysRevC.76.021901}{{\em Phys.Rev.} {\bf C76}
  (2007)  021901},
\href{http://arxiv.org/abs/0704.1647}{{\tt arXiv:0704.1647 [hep-ph]}}.

\bibitem{Denicol:2011fa}
G.~S. Denicol, J.~Noronha, H.~Niemi, and D.~H. Rischke, ``{Origin of the
  Relaxation Time in Dissipative Fluid Dynamics},''
  \href{http://dx.doi.org/10.1103/PhysRevD.83.074019}{{\em Phys.Rev.} {\bf D83}
  (2011)  074019},
\href{http://arxiv.org/abs/1102.4780}{{\tt arXiv:1102.4780 [hep-th]}}.

\bibitem{Schenke:2006yp}
B.~Schenke and M.~Strickland, ``{Photon production from an anisotropic
  quark-gluon plasma},''
  \href{http://dx.doi.org/10.1103/PhysRevD.76.025023}{{\em Phys.Rev.} {\bf D76}
  (2007)  025023},
\href{http://arxiv.org/abs/hep-ph/0611332}{{\tt arXiv:hep-ph/0611332
  [hep-ph]}}.

\bibitem{McLerran:2014hza}
L.~McLerran and B.~Schenke, ``{The Glasma, Photons and the Implications of
  Anisotropy},''
\href{http://arxiv.org/abs/1403.7462}{{\tt arXiv:1403.7462 [hep-ph]}}.

\bibitem{Martinez:2008di}
M.~Martinez and M.~Strickland, ``{Pre-equilibrium dilepton production from an
  anisotropic quark-gluon plasma},''
  \href{http://dx.doi.org/10.1103/PhysRevC.78.034917}{{\em Phys.Rev.} {\bf C78}
  (2008)  034917},
\href{http://arxiv.org/abs/0805.4552}{{\tt arXiv:0805.4552 [hep-ph]}}.

\bibitem{Shuryak:2012nf}
E.~Shuryak, ``{Monitoring parton equilibration in heavy ion collisions via
  dilepton polarization},''
\href{http://arxiv.org/abs/1203.1012}{{\tt arXiv:1203.1012 [nucl-th]}}.

\bibitem{Gelis:2013rba}
T.~Epelbaum and F.~Gelis, ``{Pressure isotropization in high energy heavy ion
  collisions},'' \href{http://dx.doi.org/10.1103/PhysRevLett.111.232301}{{\em
  Phys.Rev.Lett.} {\bf 111} (2013)  232301},
\href{http://arxiv.org/abs/1307.2214}{{\tt arXiv:1307.2214 [hep-ph]}}.

\bibitem{Berges:2013fga}
J.~Berges, K.~Boguslavski, S.~Schlichting, and R.~Venugopalan, ``{Universal
  attractor in a highly occupied non-Abelian plasma},''
  \href{http://dx.doi.org/10.1103/PhysRevD.89.114007}{{\em Phys.Rev.} {\bf D89}
  (2014)  114007},
\href{http://arxiv.org/abs/1311.3005}{{\tt arXiv:1311.3005 [hep-ph]}}.

\end{thebibliography}\endgroup
\bibliographystyle{utphys}
\end{document}